\documentclass[aps,prl,twocolumn]{revtex4}
\usepackage{graphicx,amsmath}

\newcommand{\Pe}{\mathrm{P\hspace{-1pt}e}}

\newcommand{\re}{\mbox{Re\hspace{1pt}}}

\begin{document}
\title{ Interfacial dynamics in transport-limited dissolution }

\author{Martin Z. Bazant}
\affiliation{
Department of Mathematics,
  Massachusetts Institute of Technology,
  Cambridge, MA 02139}

\date{\today}

\begin{abstract}
  Various model problems of ``transport-limited dissolution'' in two
  dimensions are analyzed using time-dependent conformal maps.  For
  diffusion-limited dissolution (reverse Laplacian growth), several
  exact solutions are discussed for the smoothing of corrugated
  surfaces, including the continuous analogs of ``internal
  diffusion-limited aggregation'' and ``diffusion-limited erosion''. A
  class of non-Laplacian, transport-limited dissolution processes are
  also considered, which raise the general question of when and where
  a finite solid will disappear. In a case of dissolution by
  advection-diffusion, a tilted ellipse maintains its shape during
  collapse, as its center of mass drifts obliquely away from the
  background fluid flow, but other initial shapes have more
  complicated dynamics.
\end{abstract}

\maketitle

The analysis of interfacial dynamics using conformal maps (``Loewner
chains'') is a classical subject, which is finding unexpected 
applications in physics~\cite{gruzberg04,bauer04,handbook05}. For
dynamics controlled by Laplacian fields, there is a vast literature on
continuous models of viscous fingering~\cite{bensimon86,howison92},
and stochastic models of diffusion-limited aggregation
(DLA)~\cite{hastings98,handbook05} and fractal curves in critical
phenomena~\cite{gruzberg04,bauer04}.  Conformal-map dynamics has also
been formulated for a class of non-Laplacian growth phenomena of both
types~\cite{bazant03}, driven by conformally invariant transport
processes~\cite{bazant04}. For growth limited by advection-diffusion
in a potential flow, the connection between continuous and stochastic
growth patterns has been elucidated~\cite{david05}, and the continuous
dynamics has also been studied in cases of freezing in flowing
liquids~\cite{maksimov76,kornev88,kornev94,alimov98}.

In all of these examples, the moving interface separates a ``solid''
region, where singularities in the map reside, a ``fluid'' region,
where the driving transport processes occur and the map is
univalent. (In viscous fingering, these are the inviscid and viscous
fluid regions, respectively.) Most attention has been paid to problems
of ``transport-limited growth'', where the solid region grows into the
fluid region, since the dynamics is unstable and typically leads to
cusp singularities in finite time~\cite{meyer82,shraiman84,howison86}
(without surface tension~\cite{bensimon86}). Here, we consider various
time-reversed problems of ``transport-limited dissolution'' (TLD),
where the solid recedes from the fluid region, e.g. driven by
advection-diffusion in a potential flow. These are stable processes,
so we focus on continuous dynamics, without surface tension.

Stochastic diffusion-limited dissolution (DLD), sometimes called
``diffusion-limited erosion'' (DLE) or ``anti-DLA'', has been
simulated by allowing random walkers in the fluid to annihilate
particles of the solid upon contact
~\cite{paternson84,tang85,meakin86,krug91}, rather than aggregating as
in DLA~\cite{witten81}.  Outward radial DLE on a lattice, or
``internal DLA'' (IDLA), where the random walkers start at the origin
and cause a fluid cavity to grow in an infinite solid, has also been
studied by mathematicians, who proved that the asymptotic shape is a
sphere in any dimension~\cite{lawlor92}. 

We begin our analysis by summarizing some exact solutions for
continuous DLD, which could help to understand fluctuations and the
long-time limits of DLE and IDLA. Although these solutions are known
in somewhat different forms for Laplacian
growth~\cite{meyer82,shraiman84,howison86}, it is instructive to
summarize them prior to considering problems of TLD by
advection-diffusion, to highlight the effects of fluid flow.

\begin{figure}
\includegraphics[width=.9\linewidth]{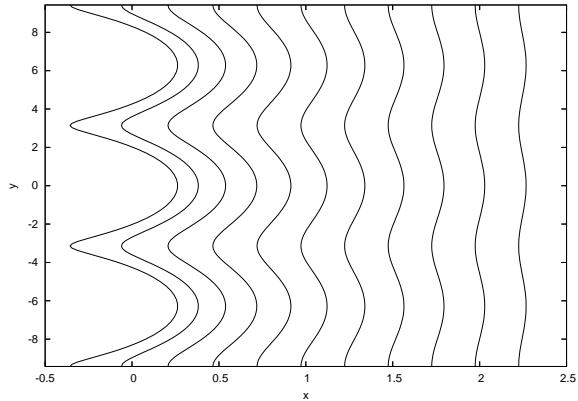}
\caption{ \label{fig:fingers} Diffusion-limited dissolution of a
  corrugated surface from left to right, for $h(0)=0$, $a(0)=.3$,
  $t=0,.25,.50,\ldots,2.50$ in Eq.~(\ref{eq:fingers}). }
\end{figure}

{\it Dissolution of surface corrugation. --} Let $G(w,t)$ be a
conformal map from the left half plane to the fluid region, where
steady (Laplacian) diffusion with uniform flux at $-\infty$ drives
dissolution of the solid. The interfacial dynamics is given by the
(dimensionless) Polubarinova-Galin (PG)
equation~\cite{handbook05,howison92},
\begin{equation}
\re\left(\overline{G^\prime}{G_t}\right) = 1, \ \ \mbox{ on } \re w = 0.
\end{equation}
An exact solution has the form,
\begin{equation}
  G(w,t) = w + h(t) + \log(1+a(t)e^w)  \label{eq:fingers}
\end{equation}
with $|a(t)|<1/2$ (real) where
\begin{eqnarray}
a &=& \frac{Ce^{-t}}{(1-a^2)^{3/2}} \sim Ce^{-t} + \frac{3}{2}C^3e^{-3t} + \ldots\\
\frac{dh}{dt} &=& \frac{1+a-a^2}{1-a^2}, \ h \sim t + C(e^{-t}-1) +
\ldots  \label{eq:ah}
\end{eqnarray}
and $C = a(0)(1-a(0)^2)^{3/2}$. This solution, shown in
Fig.~\ref{fig:fingers}, describes the decay of surface corrugations in
electropolishing~\cite{meakin86,krug91,wagner54,edwards53} or the
displacement of an inviscid fluid by an immiscible viscous fluid in a
Hele-Shaw cell~\cite{paterson84,tang85,bensimon86}. 
and closely ressembles simulations of ``mean-field DLE'' (Fig. 2 of
~\cite{tang85}).

The linear stability of a flat interface in DLD is contained in the
long-time limit of Eqs.~(\ref{eq:fingers})-(\ref{eq:ah}). With the
dimensions restored ($w \mapsto wk$, $G \mapsto Gk$, and $t \mapsto
kvt$), the interface shape becomes sinusoidal,
\begin{equation}
x(y) \sim vt + C e^{-kvt} \cos(ky),
\end{equation}
for a Fourier mode of wavenumber $k$ and mean interfacial velocity
$v$. We thus recover the classical
result~\cite{tang85,krug91,wagner54} that $kv$ is the exponential
decay rate of mode $k$, as observed in experiments~\cite{edwards53}.

\begin{figure}
\includegraphics[width=1.2\linewidth]{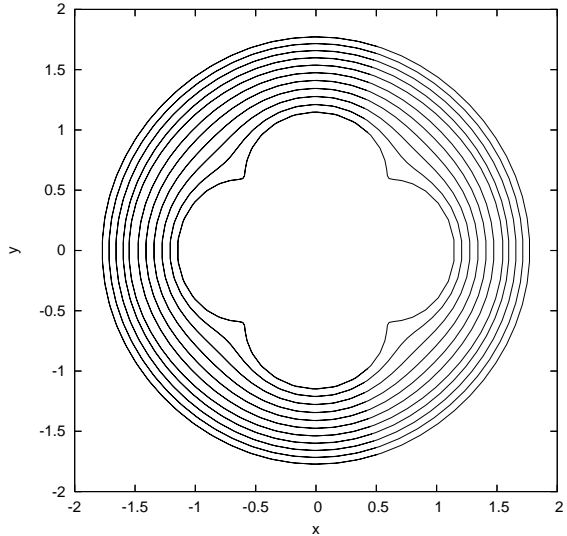}
\caption{ \label{fig:out4} Outward radial diffusion-limited dissolution
  driven by a point sink at the origin for an initial four-lobed
  perturbation of a circle. This exact solution is given by
  Eqs.~(\ref{eq:out})-(\ref{eq:outeqns}) with $N=5$, $c=0.15$, $t=0,
  0.2, 0.4,\ldots,1.0$.  }
\end{figure}

{\it Outward dissolution of clover-like shapes. -- } Next we consider
the continuous analog of IDLA: DLD in a radial geometry driven by a
constant diffusive flux from the origin. This could model quasi-steady
melting of an infinite solid around a point source of heat, or the
injection of a viscous fluid into a Hele-Shaw cell, displacing an
inviscid fluid. The radial PG equation is~\cite{handbook05,howison92},
\begin{equation}
\re\left(\overline{w g^\prime}
g_t \right) = 1, \ \ \mbox{ on }
|w|=1, \label{eq:pgr} 
\end{equation}
where $ z=g(w,t)$ is a univalent mapping of the {\it interior} of
the unit disk.
A tractable case is the clover-like $(N-1)$-fold perturbation of a
circle,
\begin{equation}
g(w,t) = a_1(t) w + a_N(t) w^N   \label{eq:gout}
\end{equation}
first analyzed by Meyer for the (time-reversed) Hele-Shaw
problem~\cite{meyer82}.  For $a_1(0)=1$ and $0 < c = a_N(0) < 1/N$
(real), the solution has the implicit form,
\begin{equation}
a_1^2 + N a_N^2 = 1 + Nc^2+2t \ \mbox{ and } \  a_1^Na_N=c,   \label{eq:aNout}
\end{equation}
Since $a_1(t)$ increases to $\infty$ from $a_1(0)=1$ and $a_N(t)$
decreases to $0$ from $a_N(0)=c<1/N < 1$, the following recursion
converges very quickly,
\begin{equation}
a_1 = \sqrt{2t + 1 + Nc^2(1-a_1^{-2N})}, \label{eq:rec}
\end{equation}
and yields asymptotic approximation upon recursive substitution.
%
An example shown in Figure \ref{fig:out4}, illustrates the rapid
smoothing of a four-leafed clover shape.

From Eqs.~(\ref{eq:aNout})-(\ref{eq:rec}) we see that the $(N-1)$-fold
radial perturbation decays as a power-law, $a_N \propto t^{-N/2}$, in
contrast to the exponential decay of perturbations of a flat
interface. We conjecture that the stochastic interface in IDLA is
asymptotic to the continuous DLD solution above, up to small
logarithmic fluctuations~\cite{lawlor95}, where $w^N$ is the smallest
perturbation in the initial cluster shape. In general, the continuous
dynamics of DLD may also accurately approximate the ensemble-averaged
stochastic dynamics of IDLA, which is not the case for DLA and other
unstable aggregation processes~\cite{david05}.

An (apparently new) explicit solution is possible for $N=2$. In that
case, Equation (\ref{eq:aNout}) reduces to a depressed
cubic~\cite{nick93}, $a_2^3 + 3p a_2 = 2 q$, solved by the
formula of Cardano and dal Ferro,
\begin{equation}
a_1 = \sqrt[3]{q + \sqrt{q^2-p^3}} + \sqrt[3]{q-\sqrt{q^2-p^3}}.
\end{equation}
where $p=-(t+c^2+1/2)/4$ and $q=-c/4$. 

\begin{figure}
\includegraphics[width=1.2\linewidth]{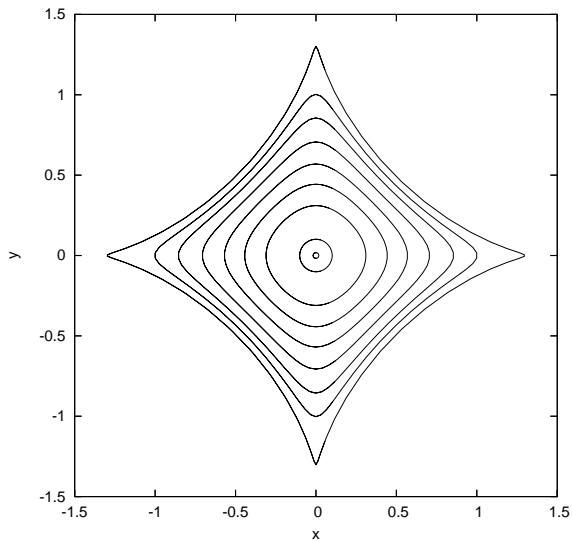}
\caption{ \label{fig:in4} Inward radial diffusion-limited dissolution
  driven by a sink at $\infty$ for a four-pointed shape.  This exact
  solution is given by Eqs.~(\ref{eq:in})-(\ref{eq:ineqs}) with $N=3$,
  $c=0.3$, $t=0, 0.04, 0.08,\ldots,0.36, 0.3649$; the collapse occurs
  at $t_c = 0.365$.  }
\end{figure}

{\it Inward dissolution of star-like shapes. -- } Next we briefly
consider inward DLD with a constant diffusive flux at infinity, which
simply is the time-reverse of Laplacian growth in a radial
geometry~\cite{howison92}. Now the map in Eq.~(\ref{eq:pgr}) must be
univalent {\it outside} the unit disk, with a Laurent expansion,
\begin{equation}
g(w,t) = \sum_{n=-\infty}^1 a_n(t) w^n, \ \ |w|\geq 1   \label{eq:ls}
\end{equation}
There are well-known solutions for $(N+1)$-fold
perturbations~\cite{shraiman84,howison86},
\begin{equation}
g(w,t) = a_1(t) w + a_{-N}(t) w^{-N} \  \ \mbox{ for } \ |w|\geq 1, \label{eq:in}
\end{equation}
where, without loss of generality, $a_1(0)=1$ and $a_{-N}(0) = c<1/N$
is real. Again, the Laurent coeficients satisfy a pair of nonlinear
equations, which is most easily solved as a fixed-point iteration,
\begin{equation}
a_1 = \sqrt{1-2t+Nc^2(a_1^{2N}-1)} \ \mbox{ and } \ a_{-N} = c a_1^N. \label{eq:ineqs}
\end{equation}
The only qualitative difference with outward DLD is that the solid
collapses to a point in a finite time, $t_c = (1-Nc^2)/2$. For $N=1$,
an ellipse ($0<c<1$) or circle ($c=0$) maintains its shape during
collapse,
\begin{equation}
g(w,t) = \sqrt{1 - \frac{t}{t_c}} \left(w + \frac{c}{w}\right)
\end{equation}
For $N\geq 1$, the shape approaches a circle prior to collapse, 
according to the asymptotic formula,
\begin{equation}
a_1(t) \sim \sqrt{2(t_c-t) + Nc^2(2(t_c-t))^N} 
\end{equation}
with $a_{-N}(t)=c a_1(t)^N$.  The collapse of a four-pointed 
shape ($N=3$) is shown in Figure~\ref{fig:in4}.

{\it Advection-diffusion-limited dissolution. --} The dynamics of
dissolution become more interesting when driven by non-Laplacian (but
conformally invariant~\cite{bazant04}) transport
processes~\cite{bazant03}, where right-hand side of the PG equation
(\ref{eq:pgr}) is replaced by the nonuniform, time-dependent flux to
the interface, $\sigma(w,t)$.  An important example is dissolution by
advection-diffusion in a potential flow, e.g. the erosion of rock by
flowing water, the evaporation of a fiber coating in a flowing gas, or
the melting of a solid column in a flowing liquid. The time-reversed
growth problem has been studied extensively in the contexts of
freezing~\cite{maksimov76,kornev88,kornev94,alimov98} and
advection-diffusion-limited aggregation~\cite{bazant03,david05}, but
it seems that dissolution -- which leads to collapse in finite time --
has not been analyzed. For a given initial shape and background flow,
{\it when and where will collapse occur?}

Consider a finite solid of constant concentration and arbitrary shape
in a two dimensional potential flow of zero concentration and uniform
velocity far away. The relative importance of advection to diffusion
is measured by the P\'eclet number, $Pe_0 = UL/D$, for a background
fluid velocity $U$, diffusivity $D$ and length $L$. The time-dependent
P\'eclet number, $Pe(t) = Pe_0 a_1(t)$, is defined by the conformal
radius, $a_1(t)$, in Eq.~(\ref{eq:ls}). When solid dissolves
($a_1(t)\to 0$), diffusion eventually dominates, and it is natural to
focus on the low-$Pe$ limit.

The flux profile $\sigma(\theta,\Pe)$ on the absorber has been studied
extensively, and very accurate asymptotic approximations are
available~\cite{choi05}. (A numerical code in matlab is also at
http://advection-diffusion.net.) From the low-$\Pe$ approximation,
\begin{equation}
\sigma \sim \frac{I_0(Pe)}{K_0(Pe/2)}e^{Pe\,\cos\theta} -
\Pe\left(\cos\theta + \int_0^{Pe} dt e^{t\cos\theta} \frac{I_1(t)}{t}\right)  \label{eq:sig}
\end{equation}
which is uniformly accurate in angle $\theta$ up to $Pe=10^{-1}$, let
us keep only the leading terms,
\begin{equation}
  \sigma \sim
  \frac{1+Pe\cos\theta}{-\log(Pe/4)-\gamma}-Pe\cos\theta \label{eq:sigsim} 
\end{equation}
where $\gamma = 0.577215\ldots$ is Euler's constant. 

In the final stage of collapse where $-\log Pe \gg \gamma$, the interface
is asymptotically circular, $g(w,t) \sim a_1(t) w$. From
(\ref{eq:sigsim}), the radius in this regime satisfies, $a_1 da_1/dt
\sim -1/\log a_1$, and thus has the asymptotic form,
\begin{equation}
a_1 \sim \sqrt{\frac{4(t-t_c)}{\log\frac{1}{t-t_c}-\log\log
    \frac{1}{t-t_c} + \ldots }}
\end{equation}

To describe the dynamics starting at small $Pe$ and ending just prior
to collapse, it is reasonable to further set $\log Pe\approx$
constant.  In this intermediate regime, the interfacial dynamics is
given by
\begin{equation}
  \re(\overline{w g^\prime} g_t ) = -1 + b\, a_1(t) \cos\theta \ \
  \mbox{ for } \ w=e^{i\theta}   \label{eq:pgb}
\end{equation}
(with a suitable choice of units). The $\cos\theta$ term has the
effect of exciting new modes in the shape of the dissolving
solid, which are not
present in the initial condition. We will see that a circle, $g(w,0) =
w$, translates away from the flow, $g(w,t) = w + a_0(t)$, as its
upstream side dissolves more quickly prior to collapse. It can also be
shown that an $(N+1)-$fold perturbation of a circle (\ref{eq:in}) will
translate and develop an $N-$fold perturbation, $g(w,t) =
a_1(t)w+a_0(t) + a_{-N+1}/w^{N-1} + a_N(t)/w^N$. Higher Fourier modes
in $\sigma(w,t)$ excite additional terms in the Laurent
expansion of the shape, $g(w,t)$.

\begin{figure}
\includegraphics[width=1.2\linewidth]{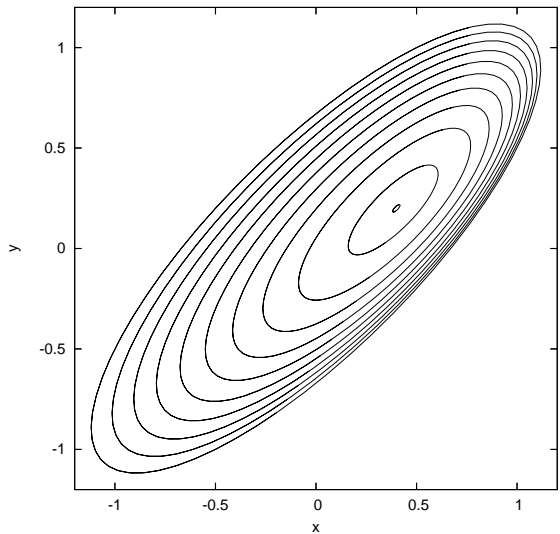}
\caption{ \label{fig:inflow} Inward advection-diffusion-limited
  dissolution of a tilted elliptical solid in a uniform background
  flow from left to right ($c=0.5i$, $b = 0.8$, $t=0, 0.04,
  0.08,\ldots,0.36, 0.3749$).  The major axis of the ellipse remains
  oriented at $\pi/4$, while its center of mass moves at an angle of
  $\pi/6$ relative to the background flow.  At time $t_c = 0.375$, the
  solid disappears at the point $a_0(t_c) = 0.4+0.2i$.  }
\end{figure}

{\it Collapse of a tilted ellipse in a uniform flow. --} To illustrate
these principles, let us consider a solid ellipse, $g(w,0) = w + c/w$
($c<1$), which is tilted at an angle $\phi=(\mbox{arg} c)/2$ with
respect to a background flow in the positive $x$ direction. During
dissolution, the shape remains an ellipse, but its center of mass,
$a_0(t)$, move. An example is shown in Fig.~\ref{fig:inflow}.

The Laurent expansion is given by
$g(w,t) = a_1(t)w + a_0(t) + a_{-1}(t)/w$,  
where $a_1(t)$ is real, but $a_0(t)$ and $a_{-1}(t)$, are
complex. Substituting into (\ref{eq:pgb}) and integrating shows that
the conformal has a square-root singularity,
\begin{equation}
a_1(t) = \sqrt{1 - \frac{t}{t_c}}, \ \ t_c = \frac{1-|c|^2}{2},
\end{equation}
since the area decreases linearly to zero (in the approximation of
constant total flux): $A(t) = A(0)-2\pi t$, where $A(0) =
\pi(1-|c|^2)$. The collapse time depends on the initial shape through
$|c|$. The slowest collapse, $t_c=1/2$, occurs for a circle, $c=0$,
while the collapse time tends to zero for a very elongated ellipse,
$|c| \to 1$, regardless of orientation.  

For a constant total flux, the collapse time $t_c$ does not depend on
the bias introduced by the flow velocity (through $b$), although the
flow affects the time scale through the initial P\'eclet number,
$Pe_0$. Mathematically, this is a general consequence of the conformal
invariance of the transport process~\cite{bazant03}, which causes the
total flux (Nusselt number) to depend only on the conformal radius and
not the asymmetric shape of the particle~\cite{choi05}. Physically,
the enhancement of dissolution on the upstream side of the solid is
cancelled by the reduction in dissolution on the downstream side.

During dissolution, the ellipse keeps its shape and its orientation
with respect to the flow direction since $a_{-1}(t) = c
a_1(t)$. However, the center of mass moves away from the flow, but
also away from the end of the ellipse which protrudes upstream. The
velocity of the center of mass is constant and in the $1+c$ direction,
\begin{equation}
a_0(t) = \frac{b(1+c)t}{1-|c|^2}.
\end{equation}
Note that the center of mass does not move along the major axis of the
ellipse at angle $\phi$, but instead at an oblique angle $\theta$
given by $\sin \theta = |c| \sin(2\phi-\theta)$. The final collapse
occurs at the point, $a_0(t_c) = b(1+c)/2$. In the simplest case,
$c=0$, a circle maintains its shape while its center of mass
translates away from the flow at (dimensionless) velocity $b$ until
collapse occurs at $x=b/2$ at time $t_c=b/2$. 

For any initial shape, the center of mass will drift away from the
flow, as well as away from any protrusions in the direction of the
flow. For non-elliptical shapes, it is nontrivial to predict the exact
time and place of collapse, even for the simplified dynamics of
Eq.~(\ref{eq:pgb}). For more general transport-limited
dynamics~\cite{bazant03}, predicting the collapse seems like an
interesting open problem.

The author thanks J. Propp for an introduction to IDLA.


\begin{thebibliography}{99}

\bibitem{gruzberg04} I. Gruzberg and L. P. Kadanoff,
  J. Stat. Phys. {\bf 114}, 1183 (2004).

\bibitem{bauer04} M. Bauer and D. Bernard, cond-mat/0412372.

\bibitem{handbook05} M. Z. Bazant and D. Crowdy, in {\it Handbook of
    Materials Modeling}, Vol. I, ed. by S. Yip, Art. 4.10 (Springer,
  2005).

\bibitem{bensimon86} D. Bensimon, L. P. Kadanoff, S. Liang,
  B. I. Shraimain, and C. Tang, Rev. Mod. Phys. {\bf 58}, 977 (1986).

\bibitem{howison92} S. D. Howsion, Euro. J. Appl. Math. {\bf 3}, 209 (1992).

\bibitem{hastings98} M. Hastings and L. Levitov, Physica D {\bf 116},
  244 (1998).

\bibitem{bazant03} M. Z. Bazant, J. Choi, and B. Davidovitch,
  Phys. Rev. Lett. {\bf 91}, 04503 (2003).

\bibitem{bazant04} M. Z. Bazant, Proc. Roy. Soc. Lond. A {\bf 460},
  1433 (2004).

\bibitem{david05} B. Davidovitch, J. Choi, and M. Z. Bazant,
  Phys. Rev. Lett. {\bf 95}, 075504 (2005).

\bibitem{maksimov76} V. A. Maksimov, Prikl. Mat. Mekh. {\bf 40}, 264 (1976).

\bibitem{kornev88} K. G. Kornev and V. A. Chugunov,
  Prikl. Mat. Mekh. {\bf 52}, 773 (1988).

\bibitem{kornev94} K. Kornev and G. Mukhamadullina,
  Proc. Roy. Soc. Lond. A {\bf 447}, 281 (1994).

\bibitem{alimov98}  M. Alimov, K. Kornev, and G. Mukhamadullina, SIAM
  J. Appl. Math. {\bf 59}, 387 (1998).

\bibitem{meyer82} G. H. Meyer, in {\it Numerical Treatment of Free
    Boundary Value Problems}, p. 202 (Birkh\"auser, 1982).

\bibitem{shraiman84} B. I. Shraiman and D. Bensimon, Phys. Rev. A {\bf 30},
  2840 (1984).

\bibitem{howison86} S. D. Howison, SIAM J. Appl. Math. {\bf 46}, 20
  (1986). 

\bibitem{paternson84} L. Paterson, Phys. Rev. Lett. {\bf 52}, 1621
  (1984).

\bibitem{witten81} T. Witten and L. M. Sander, Phys. Rev. Lett. {\bf
  47}, 1400 (1981).

\bibitem{tang85} C. Tang, Phys. Rev. A {\bf 31}, 1977 (1985).

\bibitem{meakin86} P. Meakin and J. M. Deutch, J. Chem Phys. {\bf 85}, 2320
  (1986).

\bibitem{krug91} J. Krug and P. Meakin, Phys. Rev. Lett. {\bf 66}, 703 (1991).

\bibitem{lawlor92} G. Lawlor, M. Bramson, and D. Griffeath,
  Ann. Prob. {\bf 20}, 2117 (1992).

\bibitem{wagner54} C. Wagner, J. Electrochem. Soc. {\bf 101}, 225
  (1954).

\bibitem{edwards53} J. Edwards, J. Electrochem. Soc. {\bf 100}, 189C,
  223C (1953).

\bibitem{lawlor95} G. F. Lawlor, Ann. Probab. {\bf 23}, 71 (1995).

\bibitem{nick93} R. W. D. Nickalls, Math. Gazette {\bf 77}, 354 (1993).

\bibitem{choi05} J. Choi, D. Margetis, T. M. Squires, and
  M. Z. Bazant, J. Fluid Mech. {\bf 536}, 155 (2005).

\end{thebibliography}
\end{document}